\documentclass[aps,prl,reprint]{revtex4-1}

\usepackage{graphicx}
\usepackage{dcolumn}
\usepackage{bm}
\usepackage{textcomp}
\usepackage{xcolor}
\usepackage{siunitx}
\usepackage{tikz}
\usepackage{placeins}

\begin{document}

\title{Near-monochromatic tuneable cryogenic niobium electron field emitter}

\author{C.W. Johnson$^1$}
\author{A.K. Schmid$^1$}
\author{M. Mankos$^2$}
\author{R. R\"{o}pke$^3$}
\author{N. Kerker$^3$}
\author{E.K. Wong$^1$}
\author{D.F. Ogletree$^1$}
\author{A.M. Minor$^{1,4}$}
\author{A. Stibor$^{1,2,3*}$}
\affiliation{$^1$Lawrence Berkeley National Lab, Molecular Foundry, Berkeley, USA}
\affiliation{$^2$Electron Optica Inc., Palo Alto, USA}
\affiliation{$^3$Institute of Physics and LISA$^+$, University of T\"{u}bingen, T\"{u}bingen, Germany,}
\affiliation{$^4$Department of Materials Science and Engineering, University of California, Berkeley, USA}
\email{astibor@lbl.gov}

\begin{abstract}
Creating, manipulating, and detecting coherent electrons is at the heart of future quantum microscopy and spectroscopy technologies. Leveraging and specifically altering the quantum features of an electron beam source at low temperatures
can enhance its emission properties. Here, we describe electron field emission from a monocrystalline, superconducting niobium nanotip at a temperature of 5.9 K. The emitted electron energy spectrum reveals an ultra-narrow distribution down to 16 meV due to tunable resonant tunneling field emission via localized band states at a nano-protrusion's apex and a cut-off at the sharp low-temperature Fermi-edge. This is an order of magnitude lower than for conventional field emission electron sources. The self-focusing geometry of the tip leads to emission in an angle of 3.7$^\circ$, a reduced brightness of 3.8 $\times$ 10$^8$ A/(m$^2$ sr V), and a stability of hours at 4.1 nA beam current and 69 meV energy width. This source will decrease the impact of lens aberration and enable new modes in low-energy electron microscopy, electron energy loss spectroscopy, and high-resolution vibrational spectroscopy.
\end{abstract}

\maketitle 

With the rise of novel instruments in electron  microscopy \cite{kruit_designs_2016,turner_interaction-free_2021}, spectroscopy \cite{kuster_correlating_2021,krivanek2014vibrational}, quantum electronics and electron wave function engineering \cite{madan_quantum_2020}, the demand for coherent beam sources becomes increasingly relevant in various fields of physics. Significant progress has been made in the generation of novel electron field emitters, that are e.g.~laser-triggered \cite{ehberger_highly_2015}, have a single atom apex \cite{kuo_noble_2006,pooch_coherent_2018} or consist of a LaB$_6$ nanowire \cite{zhang2022high}. The key performance goals are high brightness combined with a narrow field emission electron energy distribution, resulting in large longitudinal and transversal coherence. Such sources allow new techniques for electron phase modulation \cite{johnson_exact_2020,feist_high-purity_2020}, beam guiding \cite{hammer_microwave_2015}, interferometry \cite{schutz_biprism_2014,johnson_scanning_2021}, quantum decoherence measurement \cite{kerker_quantum_2020,beierle_experimental_2018}, novel microscopy diffraction modes \cite{latychevskaia_three-dimensional_2017}, quantum sensing \cite{pooch_compact_2017}, or even secure information transfer \cite{ropke_data_2021}. They lead to key applications in superconducting scanning tunneling microscopy \cite{huang_tunnelling_2020}, quantum electron microscopy \cite{kruit_designs_2016,turner_interaction-free_2021} or scanning Josephson spectroscopy \cite{kuster_correlating_2021}.
According to the Fowler-Nordheim (FN) description \cite{forbes_improved_2013}, the energy distribution of a beam gets more narrow with lower field emitter temperature. But the effect is negligible below room temperature and usually reported energy widths of cold field emitter are $\sim$\SI{200}{meV} \cite{peter_w_hawkes_principles_2017}. 

\begin{figure*}[t]
\includegraphics[width=1.0\textwidth]{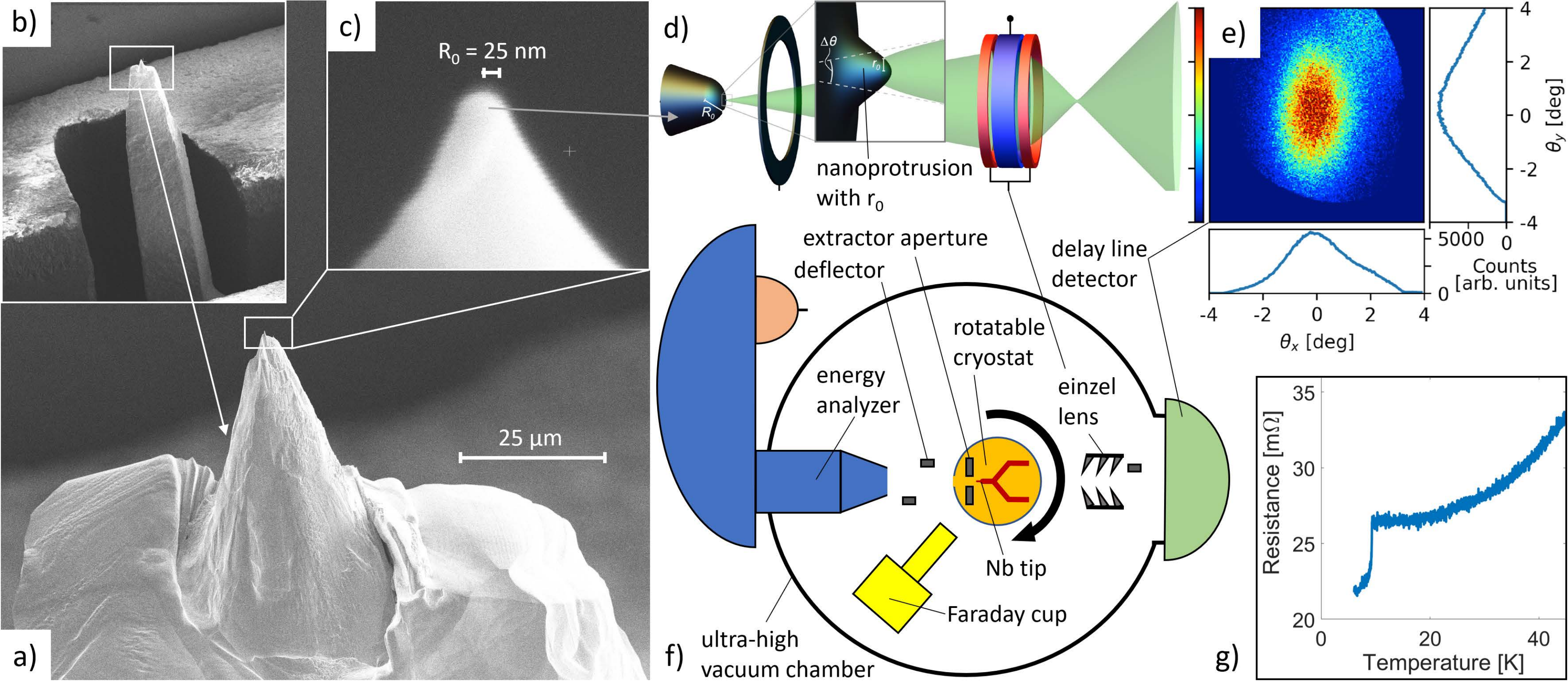}
\caption{a-c) SEM image of the FIB milled monocrystalline tip in three different magnifications. d) On top of the R$_0=$~\SI{25}{nm} radius Nb tip, an  r$_0=$~\SI{1.5}{nm}-nano-protrusion is formed that is not resolved by SEM. The nano-protrusion's field emission is magnified by an einzel lens. e) Delay line detector image of the beam's angular distribution with $V_{tip}=$ \SI{-364}{V} ($V_{ext}=$ \SI{0}{V}), yielding a mean angular spread of $\Delta\theta\approx$ \SI{3.2}{\degree}. f) Sketch of our electron field emitter characterization setup. The tip and the bar are sandwiched between two sapphire plates (visible in b) and mounted onto a rotatable cryostat perpendicular to the rotation axis, allowing it to be pointed at multiple measurement devices. g) Four-point resistance vs.~temperature measurement of the tip's bar, indicating superconducting emitter conditions.}
\label{fig:figure1}
\end{figure*}

In this Letter, we analyze the field emission properties of a monocrystalline niobium (Nb) nanotip at a temperature of \SI{5.9}{K}, well below the superconducting transition at $T_c = \SI{9.3}{K}$. A scanning electron microscope image of the tip can be seen in Fig.~\ref{fig:figure1} a-c). It is demonstrated that a surface nano-protrusion (NP) can be generated on the tip, leading to a localized quantum band state at the apex \cite{binh_field-emission_1992,gohda_total_2001}. The nano-protrusion emission (NPE) is self-focusing with an emission angle of \SI{3.2}{\degree} due to the high field enhancement at the NP apex with a small radius of curvature. The center of the NPE total energy distribution spectral peak can be tuned relative to the Fermi-energy ($E_F$) and cut-off by the sharp low-temperature Fermi-edge by changing the electric field at the emitter. It leads to an extremely narrow energy distribution full-width-half-maximum (FWHM) down to $\Delta E = \SI{16}{meV}$. This is the first monocrystalline Nb tip characterized at superconducting temperatures and the smallest reported electron field emitter energy width. It is 10-20 times narrower than the best emitters available in microscopy, leading to high coherence, while showing long-time stability and high brightness. This will lead to a paradigm shift in electron microscopy and spectroscopy, enabling electron energy loss spectroscopy (EELS) with $<$\SI{2.5}{meV} resolution and decreasing the impact of chromatic aberrations.

The Nb tip and the NP on its apex were fabricated in five steps. First, a rectangular monocrystalline wire being the base of the tip was cut out from a larger [100] oriented Nb crystal in pieces of 250$\times$\SI{250}{\micro\metre} cross-section with electric discharge machining \cite{surface}. Then this wire was spot-welded in the center of a V-shaped polycrystalline Nb wire bar. As a next step, the monocrystalline tip base was electrochemically etched down to a length of only $\sim$\SI{2}{mm} in a KOH solution according to a procedure described in \cite{hou_nanoemitter_2016}. This led to the $\sim$\SI{100}{\micro\metre}-base shown in the SEM images of Fig.~\ref{fig:figure1} a-b). It is then further shaped by gallium ion beam milling in a FIB \cite{Helios} to fabricate the conical tip in Fig.~\ref{fig:figure1} c) with a radius $R_0=$25$\pm$\SI{2}{\nano\metre}. The most delicate step is the formation of a NP (\SI{<5}{nm} \cite{binh_field-emission_1992}) on the tip after being installed and cooled on the cryostat. This is performed through several annealing cycles by ramping up a current of \SI{4.85}{A} through the polycrystalline Nb wire bar until the tip is glowing at $\sim$950 $^\circ$C \cite{Leeds} and simultaneously setting a \SI{-3}{kV} electrical bias on the extractor. By automating this annealing process with a programmable current source, NP formation on our tip is reproducible and the NPE is stable on the order of hours. However, there is still variability in the emission properties of the NPs after each annealing process, making several annealings necessary to get the optimal conditions. 
Our NP survives at a minimum \SI{82}{K} \cite{Johnson2022b}. We assume it to be RT-stable and made of niobium due to corresponding literature data for 
tungsten and gold tip NPs fabricated by a similar field surface melting technique \cite{binh_electron_1992}. This is in contrast to diffusion-growth NPs such as in \cite{nagaoka_field_2001} which are believed to be composed of contaminant molecules.

The setup to cool the tip and characterize its field emission properties is illustrated in Fig.~\ref{fig:figure1} f). It allows a high-resolution measurement of the beam energy spectra, the emission current, and the angular distribution. The nanotip and the extractor are cooled by a closed-cycle liquid helium cryostat \cite{Advanced} and thermally isolated by a copper cooling shield with a small opening for the beam path. The step in the four-point resistance measurement of the V-shaped Nb-bar in Fig.~\ref{fig:figure1} g) is used to calibrate the measured temperatures to $T_c$ and reveals that the emitter is operating under superconducting conditions at \SI{5.9}{K}. The cryostat is mounted inside an ultra-high vacuum chamber with a pressure of $\sim$\SI{8e-11}{Torr} on a rotational flange and a 3D-manipulator. It allows positioning and aligning of the emitter towards three measurement units. The first one is a hemispherical electron beam energy analyzer \cite{ScientaOmicron} with a resolution of \SI{3}{meV}. Between the analyzer and the tip, a double deflector is installed for beam alignment. A rotation of the cryostat points the tip towards a Faraday-cup \cite{Kimball} with a pico-ammeter \cite{Keithley6430} for the beam current measurements. Another rotation points the emitter towards a microchannel plate delay line detector (DLD \cite{Roentdek}) with a single-electron resolution to measure the beam's angular distribution. A custom einzel lens with a deflector is also installed at this position for magnification and alignment. The analyzer, Faraday cup, and DLD operate in different regimes of current and tip voltage, allowing a wide range of beam analysis. This realizes a unique system that combines spectroscopy, beam current, spatial distribution, and temperature control.

\begin{figure*}[t]
\begin{tikzpicture}
    \node[scale=1.04,inner sep=0pt] at (0,0) {\includegraphics[width=0.95\textwidth]{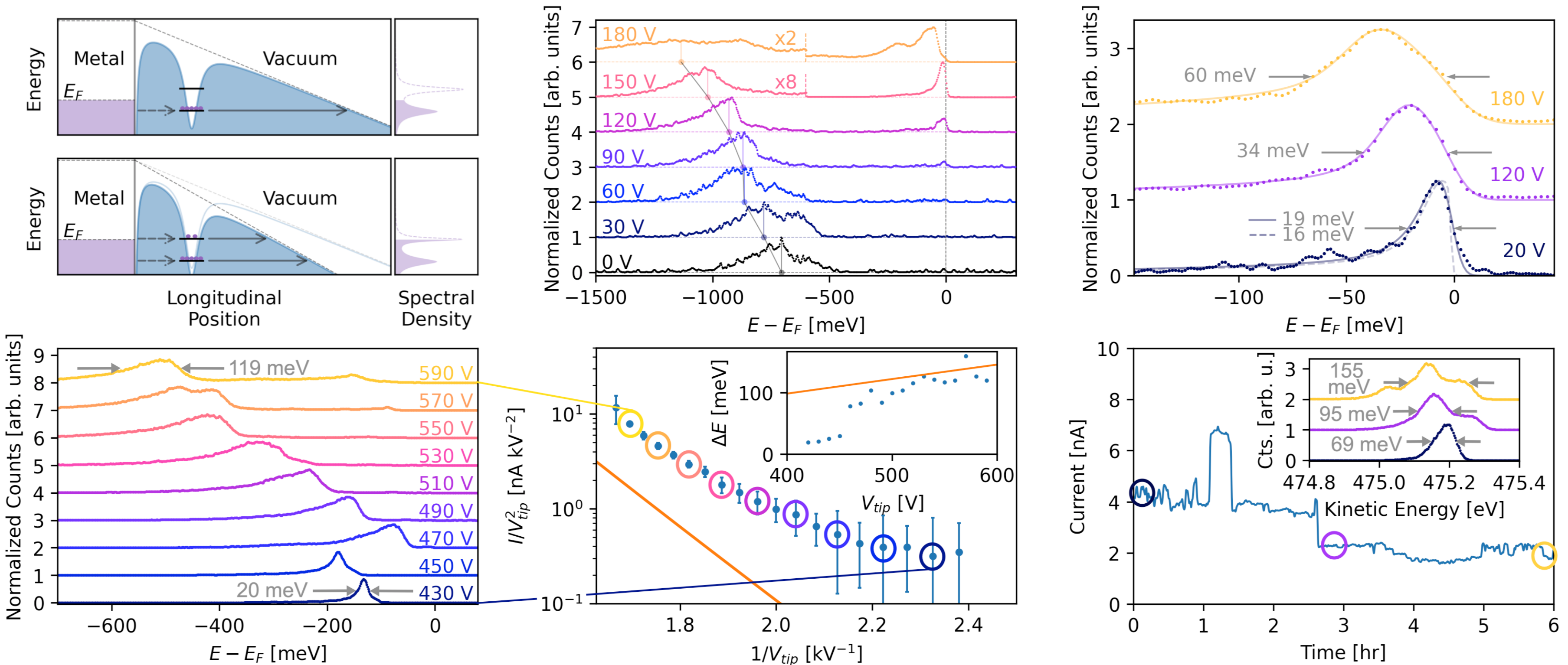}};
    \node[scale=1,inner sep=0pt] at (-9.0,3.25) {a)};
    \node[scale=1,inner sep=0pt] at (-3.04,3.25) {b)};
    \node[scale=1,inner sep=0pt] at (2.85,3.25) {c)};
    \node[scale=1,inner sep=0pt] at (-9.0,-0.3)  {d)};
    \node[scale=1,inner sep=0pt] at (-3.04,-0.3) {e)};
    \node[scale=1,inner sep=0pt] at (2.85,-0.3)
    {f)};
\end{tikzpicture}
\caption{ a) Qualitative illustration showing NP's resonant states in the Coulomb barrier, allowing electrons an enhanced field emission to the vacuum. Lower panel: Decreasing the field strength in resonant tunneling emission can shift resonances originally above E$_F$ into the occupied region, resulting in a tuneable sharp cut-off at low temperatures and an ultra-narrow beam energy distribution. b) Experimental spectra showing the nano-protrusion (NP) emission peaks decreasing in energy by increasing the extractor voltage $V_{ext}$ from 0 to \SI{180}{V} at a fixed tip voltage of $V_{tip}=$ \SI{-475}{V}. c) The narrowest energy distribution observed of \SI{16}{meV} after noise correction (\SI{19}{meV} before). The small energy width is due to the use of the sharp low-temperature Fermi edge to increasingly truncate the higher energy portion of the NP’s discrete state emission by applying an extraction voltage of 180, 120, and \SI{20}{V}, pushing the resonance peak over E$_F$. d) Energy spectra corresponding to the color-matching encircled data points in e) with the tip voltage labeled on the right side ($V_{ext}=0$). The spectra with increasing V$_{tip}$ correspond to the circles in e) from right to left. e) Fowler-Nordheim (FN) plots of a NP’s emission current with decreasing $V_{tip}$ from \SI{-600}{V} to \SI{-420}{V}. The orange line is a theory expectation for a pure FN behavior, confirming that the resonant emission process is different compared to a FN emitter. f)~Stability measurement of the Nb field emitter over \SI{6}{hours}. The sudden increase after \SI{1}{hour} is a familiar effect observed in other cold field emitters and can be explained by the temporary presence of an adatom on the emission site. The inset shows energy spectra taken at the beginning of the recording of f), after \SI{3}{h} and at the end. A reasonable peak broadening is visible, presumably due to adatoms accumulating near the tip's apex or small structural changes that alter the boundary conditions for the NP resonant emission. The data in b)-f) are recorded after different annealing cycles, with slight variations in the nano-protrusion geometry and resonance conditions. The peak amplitudes in the spectra are normalized. }  
\label{fig:figure2}
\end{figure*}
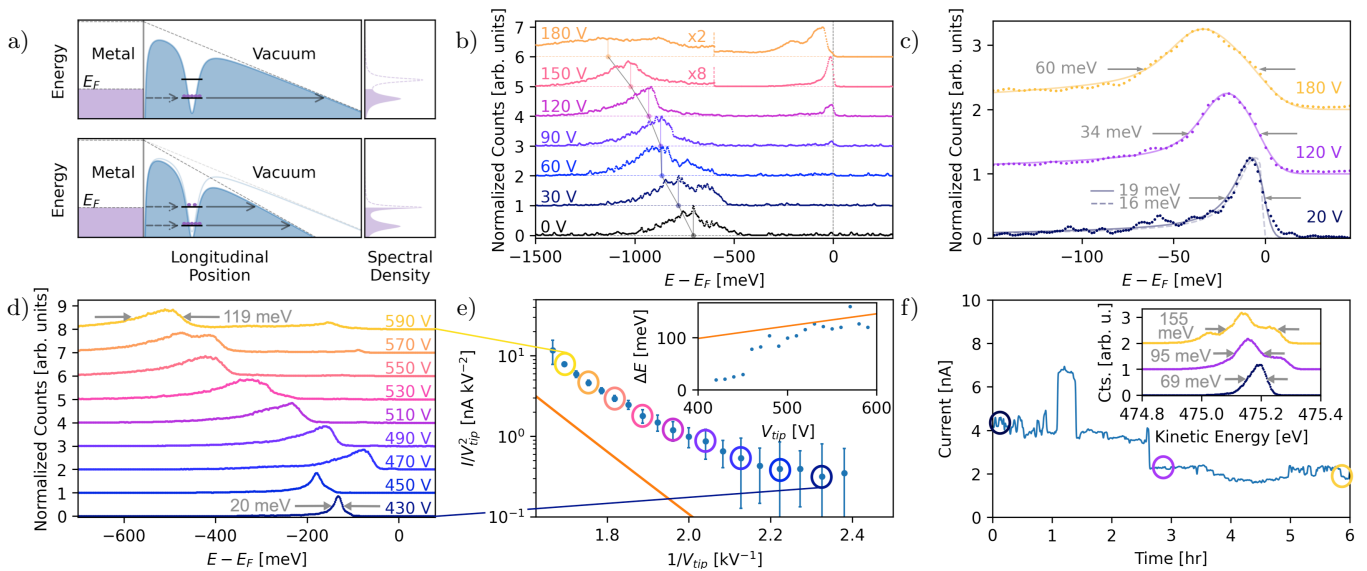

The energy spectrum of FN cold field emission of electrons from a metal's Fermi sea to vacuum with respect to $E_F$ is given by the product of the Fermi-Dirac distribution function and an exponential decay term $G(E) \propto f(E)\exp(E/D) \label{eq:FN_E}$. Here, $f(E)=(1+\exp(E/k_BT))^{-1}$ and $D=e\hbar F/2\sqrt{2m\phi}$ are the energy scale parameter for field emission with the work function $\phi$ and the field strength at the emission site $F=|\mathbf{E}|$ \cite{young_theoretical_1959}. The field strength can be determined by $F=\beta V_{tip}$, where $\beta=(kR_0)^{-1}$ with $R_0$ being the base radius and a geometry factor typically taken as $k=5.7$ for a metallic tips \cite{keramati_photofield_2020}. The FN emission current density goes as $J\propto F^2\exp(-B\phi^{3/2}/F)$, where $B\approx$ \SI{7}{V\per eV\tothe{3/2}nm}. At cryogenic temperatures, $f$ is nearly a step function, causing the energy spectrum width to scale as $\Delta E=\ln(2)D$ (with FWHM, the solution to $\exp(\Delta E/D)=1/2$) due to the decreasing tunnel width through the Coulomb barrier with increasing $F$. 

The addition of a NP with radius $r_0\ll R_0$ on the tip apex creates localized, discrete electronic states that can act as intermediate levels for resonant tunneling through the Coulomb barrier, and FN theory alone is no longer sufficient to model the emission. The energy spectrum of NPE via these resonant states can be approximated as,
$G_{tot}(E) = G(E)\left(1+\sum_nR_n(E-E_n+\alpha F)\right), \label{eq:FN_E_RT}$
where the resonant enhancement factors $R_n$ are single peaked functions (Lorentzian distributions in analytic approximation \cite{lin_field-emission_1992}) corresponding to each discrete state at energy $E_n$ that shifts linearly with the applied field strength at the tip. The maxima of the resonant enhancement factors can be $|R_n|_{max}\gg \exp(-(E_n-\alpha F)/D)$ for emission well below $E_F$, this means that NPE with isolated bright resonant tunneling spectral peaks can occur even before any FN field emission is measurable. Additionally, at cryogenic temperatures, the width of the Fermi edge is much narrower than the width of any $R_n$ peak, allowing NPE peaks to be 'turned off' or partially truncated by tuning over $E_F$. The resonant tunneling field emission process is illustrated in Fig.~\ref{fig:figure2} a). The electric field around a biased NP on a nanotip autofocuses the NPE into an emission angle $\sim$2\SI{-6}{\degree} \cite{saenz_field_1990}, as illustrated in Fig.~\ref{fig:figure1} d). This is much smaller than emission from a tip geometry without the NP which would be within $\sim$20\SI{-30}{\degree} \cite{saenz_field_1990}. Using an einzel lens with a calibrated magnification and the DLD, we determined the angular beam distribution in Fig.~\ref{fig:figure1} e) as $\Delta \theta\approx$ \SI{3.2}{\degree} (azimuthally averaged emission angle). With a reasonable assumption that the NP is hyperboloidal, we estimate the size of the emission radius on the NP from the measured values of $R_0$ and $\Delta\theta$, using the relation $r_0=R_0\phi\Delta\theta^2/2D$ \cite{saenz_field_1990}. Inserting in the appropriate values for our tip $R_0=$ 25$\pm$\SI{2}{nm}, $\Delta\theta=$ 3.2$\pm$\SI{0.25}{\degree}, $D=$ \SI{118}{meV},  $\phi=$ \SI{4.5}{eV} at $V_{tip}=$ -364$\pm$\SI{2}{V}, we find an emission site radius of $r_0=$ 1.5$\pm$\SI{0.3}{nm}.

Furthermore, we recorded the energy spectrum of the NPE with an energy analyzer. By keeping the tip voltage constant and varying the extractor voltage to tune the NPE with respect to $E_F$, we recorded the NPE energy spectra in Fig.~\ref{fig:figure2} b) from two distinct NP states. The broad lower energy peak is shifted away from $E_F$ at a linear rate of $\sim$2 \si{\milli\electronvolt\per\volt}. The higher energy peak is initially above $E_F$ and begins to cross $E_F$ at $V_{ext}=$ \SI{90}{V}. It has a $\Delta E=$ \SI{34}{meV} FWHM at $V_{ext}=$ \SI{150}{V} and completely passes $E_F$ at $V_{ext}=$ \SI{180}{V} showing a $\Delta E=$ \SI{100}{meV} FWHM main peak. Nominally similar values of peak widths and energy shifts are regularly reproduced after each tip anneal and NP formation. 

The direct tunability of these discrete NP states by the tip or extractor voltage allows us to reduce the energy width by truncating a portion of the NPE spectrum, with the sharp (less than \SI{1}{meV}), low-temperature Fermi-Dirac function at $E_F$. It is significantly narrower than the FWHM of the NPE from the discrete states, producing a total FWHM $\Delta E=$ \SI{19}{meV} at a tip voltage of $V_{tip}=$ \SI{-475}{V} with $V_{ext}=$ \SI{20}{V}. This is the smallest $\Delta E$ value ever recorded for field-emitted electrons. The result is presented in Fig.~\ref{fig:figure2} c) showing the spectrum width decreasing at lower $V_{ext}$ as the peak shifts toward $E_F$ from below. We use $G_{tot}(E)$ convolved with a kernel, being a normalized Gaussian distribution $g(E,dE)=\exp(-x^2/2dE^2)/\sqrt{2\pi dE^2}$ to account for systematic instrument noise (analyzer, voltage source, vibrations), as a fit for the total energy distribution: $G_{fit}(E) = G_{tot}(E)*g(E)$. We took $R_n(E) = r_n/((E-E_n)^2+\Gamma_n^2)$ as the form for the resonant enhancement factors, where $r_n$ are the resonance amplitudes, and $\Gamma_n$ are the resonance linewidths. With $r_n$, $E_n$, $\Gamma_n$, and $dE$ as fitting parameters, we performed a least-squares-fit to the experimental spectra. Using the fit parameters $r_n$, $E_n$, $\Gamma_n$ in the function $G_{tot}$, we determined an \SI{8.2}{meV} system energy resolution. If it is deconvolved from the measured energy distribution, our narrowest emission peak has a FWHM of \SI{16}{meV}.

In the next step, a Faraday cup was used to measure the total NPE current. After further annealing, a NP was formed and the tip voltage was decreased from $V_{tip}=$ \SI{-600}{V} to $V_{tip}=$ \SI{-420}{V} in steps of \SI{10}{V} at zero extraction voltage, and the emission current was recorded. The $I/V$-data is shown in Fig.~\ref{fig:figure2} e) in a FN representation, where the orange linear relation between $1/V_{tip}$ and $\log\left(I/V_{tip}^2\right)$ would be expected to apply the FN theory. Clearly, our measured resonance emission data indicates a strong deviation from FN. After this measurement, the tip was pointed towards the energy analyzer for the recording of the spectra at the same tip voltages. It allows correlating the current data with the energy widths of the emissions, as shown in Fig.~\ref{fig:figure2} d). The inset in Fig.~\ref{fig:figure2}~e) plots the energy FWHM $\Delta E$ as a function of the tip voltage, again with an orange line reflecting the FN expectation. It demonstrates that resonant field emission can provide a significantly narrower energy width than the FN process. After further annealing, we performed a field emission stability measurement over \SI{6}{h}, as shown in Fig.~\ref{fig:figure2} f). It was started by the recording of a spectrum at $V_{tip}= -$\SI{480}{V} with an energy FWHM of \SI{69}{meV} at an emission current of about \SI{4.1}{nA}. Subsequently, the current was recorded for \SI{3}{h} before the tip was pointed again towards the energy analyzer, where a spectrum with a FWHM of \SI{95}{meV} was recorded. After further current measurement for \SI{3}{h}, a final spectrum was taken with a width of \SI{155}{meV}. The spectra are shown in the inset of Fig.~\ref{fig:figure2} f). Finally, the angular beam distribution (at a lower tip voltage to avoid DLD damage) presented in Fig.~\ref{fig:figure1} e) was recorded. 

One important figure of merit for electron sources is the reduced brightness $B_r= J/\Delta\Omega V_{tip}$, where $J=I/\pi r_0^2$ is the current density, and $\Delta\Omega=\pi\Delta\theta^2/4$ is the solid angle of emission. This is an invariant quantity along the beam path and a more apt way to compare different emission sources than current alone. In the NP model, the angle of emission goes as $\Delta\theta\propto \sqrt{V_{tip}}$ \cite{saenz_field_1990}. Since we are unable to measure the angular distribution of the NPE at high currents with the microchannel plate of the DLD, we can use the relation $\Delta\theta_2=\Delta\theta_1\sqrt{V_{tip,2}/V_{tip,1}}$ to determine the high current, \SI{4.1}{nA}, angular distribution $\Delta\theta_2=$ 3.7$\pm$\SI{0.29}{\degree} at $V_{tip,2}=-$480$\pm$\SI{3}{V} from the measurement in Fig. \ref{fig:figure1} e). Then, using the determined NPE radius $r_0=$ 1.5$\pm$\SI{0.3}{nm}, we find a reduced brightness of our source to be $B_r=$ (3.8$\pm$2.1)$\times10^8$ \SI{}{A \per (\square m \steradian V)} at an energy FWHM of \SI{69}{meV}. 
This extremely high brightness combined with the low energy width outperforms commercial cold field emission sources \cite{borrnert_electron_2018,shao2018high} and other NPE tips \cite{nagaoka_field_2001}. For comparison, the  commercial source with a monochromator in \cite{mook2000construction} has a comparable energy distribution of \SI{61}{meV} but only a reduced brightness of \SI{6.3e6}{A \per (\square m \steradian V)}. 
Our calculations reveal that applying a monochromator to a beam with \SI{4}{nA} current at \SI{69}{meV}, such as in Fig.~\ref{fig:figure2}~f), will lead to \SI{400}{pA} with \SI{8}{meV} or \SI{100}{pA} with \SI{2.5}{meV}. This is more than an order of magnitude improvement to recent vibrational spectroscopy measurements \cite{krivanek2014vibrational}. We are working on improving the long-term stability and repeatability of the emission to make this field emitter a commercially viable electron source.

Having field emitted electrons with such a narrow energy distribution coming from a nanoscale spatial location raises the question about the ultimate limit in the width of the energy spectrum. For instance, in FN field emission at cryogenic temperatures the transverse and longitudinal uncertainties in momentum are $\sigma _{p\perp}\approx\sigma _{p\parallel}$ \cite{peter_w_hawkes_principles_2017}. As an estimate, we set the size of the emission region to be the uncertainty in the position $\sigma _x\approx r_0=$ \SI{1.5}{nm} and we approximate the uncertainty in the momentum from the energy distribution $\sigma_ p\gtrsim\sqrt{2m\Delta E}=$ \SI{4.3e-16}{eV s\per nm} of our measured $\Delta E=$ \SI{16}{meV}. Then we have $\sigma_x\sigma_p\approx 1.9\times\frac{\hbar}{2}$, with the caveat that we used a calculated estimate value of $\sigma_x$ and a partial estimate in momentum, excluding $\sigma _{p\perp}$. It reveals that the uncertainty product of our source is close to a factor of two from the Heisenberg limit.

In summary, we realized a superconducting monocrystalline niobium nanotip field emitter at a temperature of \SI{5.9}{K} with an unprecedented ultra-narrow tuneable energy distribution down to \SI{16}{meV} FWHM, a small emission angle of \SI{3.2}{\degree} and a brightness of up to \SI{3.8e8}{A \per (\square m \steradian V)}. 
This was achieved by the fabrication of a nano-protrusion at the tip apex, leading to a localized quantum state that mediates resonant tunneling through the Coulomb barrier beyond the Fowler-Nordheim model. Although the tip emission varies slightly after each preparation cycle, this low energy distribution was measured several times following tip preparation and could be reproduced with a completely new ion milled tip (see appendix). The energy width is in the same range as the recently developed flat surface source based on near-threshold photoemission from single-crystal Cu(100) at $\SI{35}{K}$ with $\Delta E = \SI{11.5}{meV}$ \cite{karkare_ultracold_2020}. However, the emission area of our source is orders of magnitude smaller, providing potential advantages in spatial coherence and brightness. It opens up new fields such as high-resolution vibrational spectroscopy in the electron microscope \cite{krivanek2014vibrational} with sub-meV resolution when combined with a monochromator. It will improve EELS enabling isotopic analysis and mapping in transmission electron microscopy \cite{hachtel2021isotope}, decrease the impact of chromatic aberrations in low-voltage scanning electron microscopes and enhance the accuracy in the study of band gaps or defects to the sub-nanometer level. The small spatial dimension of the beam emission area in the nanometer regime combined with the narrow energy distribution yields a high longitudinal and transversal coherence of this source. This will benefit emerging techniques such as multipass-transmission electron microscopy \cite{juffmann2017multi} and quantum electron microscopy \cite{kruit_designs_2016} as well as quantum information science applications with coherent electrons \cite{ropke_data_2021} or metrology with electron interferometers \cite{pooch_compact_2017}. 
\begin{acknowledgments} 
We thank Alexei Fedorov for his fruitful assistance. Work at the Molecular Foundry was supported by the Office of Science, Office of Basic Energy Sciences, of the U.S. Department of Energy under Contract No.~DE-AC02-05CH11231. This material is based upon work supported by the U.S. Department of Energy, Small Business Innovation Research (SBIR) / Small Business Technology Transfer (STTR) program under Award Number DE-SC-0019675. We also acknowledge support by the Deutsche Forschungsgemeinschaft through the research grant STI 615/3-1.
\end{acknowledgments} 

\begin{figure}[b]
\begin{tikzpicture}
    \node[scale=0.45,inner sep=0pt] at (-5.5,0) {\includegraphics[width=1\textwidth]{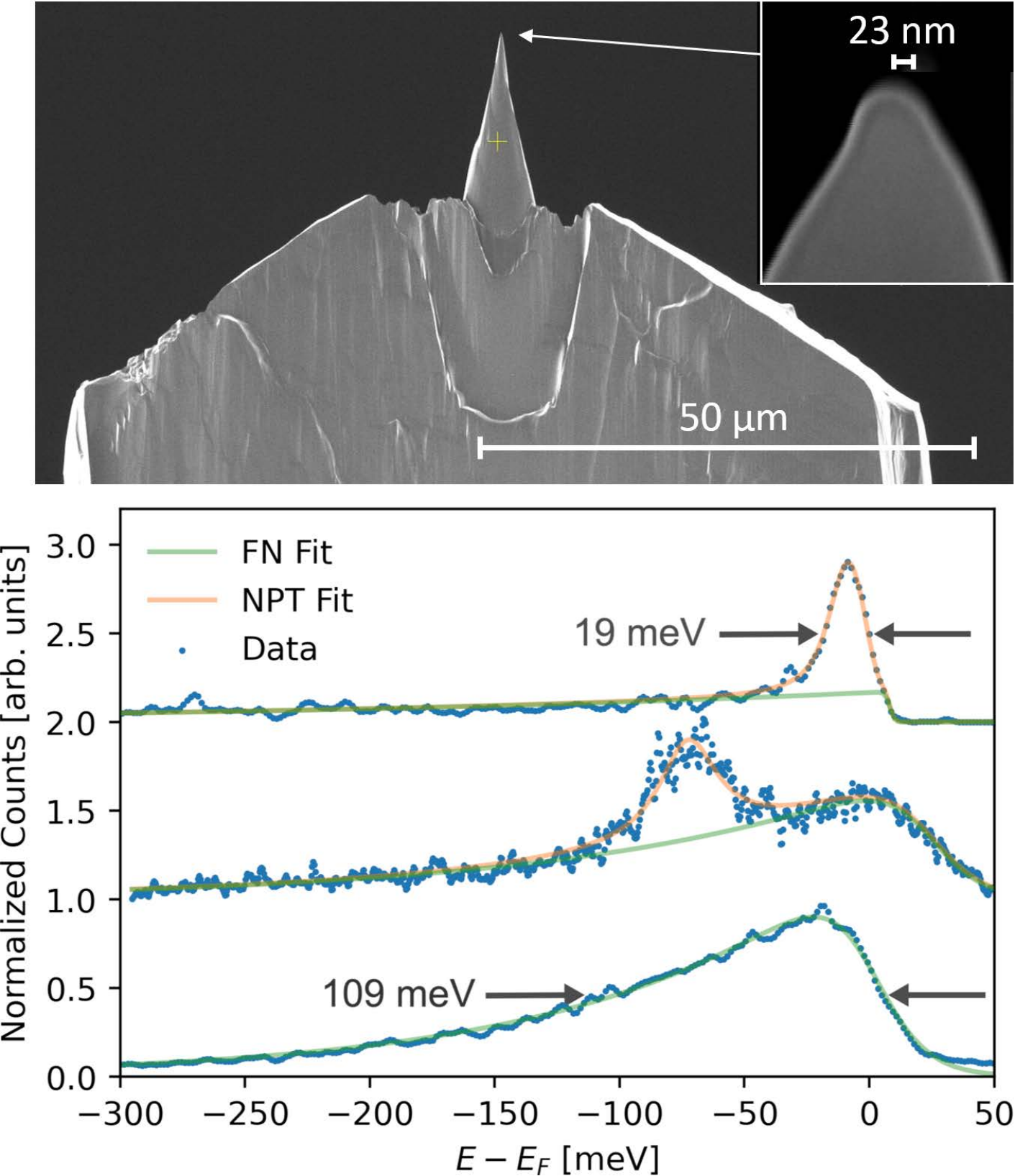}};
    \node[scale=1,inner sep=0pt] at (-9.5,1.0) {(a)};
    \node[scale=1,inner sep=0pt] at (-9.5,-4.45) {(b)};
    
\end{tikzpicture}
\caption{a) SEM image of a separately fabricated superconducting Nb tip to demonstrate reproducibility. Inset: Higher magnification, revealing a radius of \SI{23}{nm}. b) Upper curve: the energy spectra of the separate NPT. The orange line is a fit according to a NPT model, the green line fits the FN theory. The emission is mainly due to the localized band states in the NP and cut off by the sharp low-temperature Fermi-edge, decreasing the FWHM of the peak to \SI{19}{meV}. Middle curve: Mixed emission with a contribution of the band state NPE and an underlying FN distribution. Lower curve: Increasing the annealing temperature destroys the NP on the tip apex and leads to a pure FN field emission energy spectra with a FWHM of \SI{109}{meV}. } 
\label{fig:figure3}
\end{figure}
\vspace{1.5cm}
\noindent {\bf Appendix} \\
\\
Reproducibility is an important issue in the fabrication of new field emitters. Occasionally, it took several annealing cycles at the same temperature to get a narrow NPE. However, the \SI{19}{meV} FWHM distribution was usually stable for hours and could be regained repeatedly after annealing on different days and after the tip cooled from room temperature back to \SI{5.9}{K} when the cryostat was turned off and on again. We additionally reproduced the \SI{19}{meV} NPE (\SI{16}{meV} with instrument resolution correction) with a different tip, shown in Fig.~\ref{fig:figure3}~a). It was fabricated from the same monocrystalline Nb wire, but with separate etching, ion beam milling, and annealing procedures. For this tip, the annealing temperature was increased stepwise. This leads to a transition from a tip with a NP (NPT) to a bare tip without the NP. Typical energy spectra for these cases and a mixed intermediate emission are shown in Fig.~\ref{fig:figure3} b). In the upper curve, we measured the single ultra-narrow resonant NPE peak from a localized band state created by a NP that is cut off at $E_F$ as discussed in the main text. The field emission tip voltage was $V_{tip}\approx$ \SI{-650}{V}. The spectra in the middle curve shows a mixed emission with a Lorentzian peak from the NPE (indicated by the orange NPT fit curve) on top of a FN-shaped emission (indicated by the green FN fit curve), at $V_{tip}\approx$ \SI{-600}{V}. Here, the NPE peak is shifted away from $E_F$ to a lower energy by the application of an extractor voltage. It is no longer cut off by the Fermi-edge, leading to a significantly broader energy distribution. After increasing the annealing temperature, the NP disappears and the field emission starts at a higher tip voltage of \SI{-880}{V}. The resulting energy spectrum on the bottom of Fig.~\ref{fig:figure3}~b) is significantly broader and solely according to the FN theory. The ability to form a NP was diminished after annealing the tip to a certain threshold temperature, causing the base tip radius to blunt. The bottom distribution in Fig.~\ref{fig:figure3}~b) is a typical emission spectrum after this irreversible transformation process.

The narrow emission of \SI{19}{meV} from the NPT in the main text was only observed at rather low beam currents in the pA-regime, in contrast to the high currents shown in the stability measurement in Fig.~\ref{fig:figure2}~f) (nA-regime). We assume the energy deposition on the beam exit area in the NP is not negligible at higher beam currents, causing a change of the quantum state boundary conditions in the NP that leads to a higher energy width.  This is supported by observations in \cite{vu_thien_binh_local_1992} where the local beam exit area temperature changed by \SI{210}{K} between emission currents from \SI{30}{pA} to \SI{4}{nA} on a tungsten tip NP starting at room temperature. Due to the low beam heating on the electron extraction area, we assume that the stability for the \SI{19}{meV} emission is significantly longer than for the $\sim$\SI{4}{nA} in Fig.~\ref{fig:figure2}~f). However, such a measurement would require a higher resolution than available in our setup.

The emission properties of the protrusion on our tip were modeled in two steps. First, we introduced the concept of field enhancement due to tip geometry in the context of a bare tip and regular FN emission. Then, the protrusion on the tip was added by applying the analytical theories derived in \cite{lin_field-emission_1992,saenz_field_1990}, where the electric field parameter is calculated considering the ratio of the base radius and the protrusion radius to estimate the additional field enhancement.

\bibliographystyle{apsrev4-1}
\bibliography{mybib}

\end{document}